# Light refraction by water as a rationale for the Poggendorff illusion

Sergey I. Bozhevolnyi

*Centre for Nano Optics, University of Southern Denmark, Campusvej 55, DK-5230 Odense M, Denmark*

*Email: seib@iti.sdu.dk*

**Abstract**

The Poggendorff illusion in its classical form of parallel lines interrupting a transversal is viewed from the perspective of being related to the everyday experience of observing the light refraction in water. It is argued that, if one considers a transversal to be a light ray in air and the parallel lines to form an occluding strip of a medium with the refractive index being between that of air and water, then one should be able to account, both qualitatively and quantitatively, for most of the features associated with the Poggendorff illusion. Statistical treatment of the visual experiments conducted with 7 participants, each analysing 50 configurations having different intercepting angles and strip widths, resulted in the effective refractive index of the occluding strip $N = 1.13 \pm 0.15$, which is sufficiently close to the average (between that of water and air) refractive index of $\sim 1.17$. It is further argued that the same mechanism can also be employed to account for many variants of the Poggendorff illusion, including the corner-Poggendorff pattern, as well as for the Hering illusion.

# 1. Introduction

The Poggendorff illusion, which in its classical form can be described as perceived non-collinearity of two collinear segments abutting obliquely on two parallel (interrupting) lines, and its many variants have been a subject of a considerable amount of speculations and experimental investigations for several decades (for review, see Ninio, 2014). The main features, which include a strong influence of the angle between the (collinear) segments and interrupting lines, its scaling with the distance between the interrupting lines and its relative independence on colouring of the space (strip) between the interrupting lines, has been well known for a long time (Tolansky, 1964). Furthermore, for the traditional Poggendorff display described by the width $W$ between the parallels and the acute angle $\alpha$ between the segments and parallels, the misjudgement $M$ measured along the parallel (Fig. 1) has been found to follow the relation $M \cong 0.162W \cot\alpha$ (Weintraub & Krantz, 1971). Additionally, quite important and detailed observations were made with the simplified pattern composed of two parallel lines, an oblique abutting on one line, and a small disk on the other line, collinear with the oblique (Weintraub et al., 1980). It has been pointed out there and corroborated elsewhere (Ninio & O'Regan, 1999), that the main determinant of the illusion is the orientation of the oblique, with the illusion being enhanced when the angles decreased in absolute values, and that there were maxima around the −45° and +45° orientations of the oblique. Another interesting observation is related to the relatively well established fact that the illusion disappears when only the acute components are shown and re-appears when the pattern is changed to showing only obtuse ones (Weintraub & Krantz, 1971; Pressey & Sweeney, 1972; Weintraub & Tong, 1974; Day & Dickinson, 1976; Talasli & Inan, 2015).

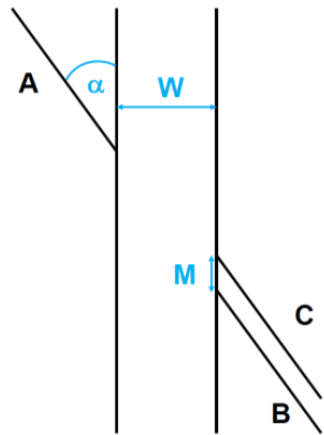

Fig. 1. Schematic of the Poggendorff illusion in its classical form as shown in a book published over 50 years ago (see Fig. 36 in Tolansky, 1964). The illusion is in the impression that it is segment C that is a continuation of line A, while it is, in fact, lines A and B that are collinear. Notations used in quantification of the illusion are also shown.

There is no shortage in theories explaining the origin(s) of the Poggendorff illusion (see Ninio, 2014; Talasli & Inan, 2015, and references therein). Some of the theories, for example, suggest that a rationale behind the illusion is based on perspective effects in a carpentered world (e.g., Gillam, 1971; Phillips, 2006), or in natural surroundings (Howe et al., 2005). Many various theories involving perceived (in one or another manner) shrinkage of occluded entities, including the most recent explanation that is based on the size-distance invariance hypothesis in its special case known as Emmert's law (Talasli & Inan, 2015). Leaving a detailed comparison of various theories existing in abundance to special reviews (e.g., Ninio, 2014), I would only like to note that a unified explanation (even qualitative, not to speak about quantitative one) of all sides of the Poggendorff illusion seems still to be missing.

In this paper, I would like to suggest another rationale, viz., everyday experience of light refraction by water, for explaining the Poggendorff illusion. In the following, I first obtain the corresponding mathematical expression, describing the perceived shift (misjudgement) M as a function of configuration parameters, as well as its simplified version that agrees well with the previously established one (Weintraub & Krantz, 1971). The main result is then corroborated with the experimental observations conducted by 7, rather different in age and background, individuals. Finally, I speculate on the features of the Poggendorff illusion and its variants that are outside of the validity domain of the proposed mechanism.

## 2. Quantitative description

Contemplating the classical Poggendorff illusion shown in Fig. 1 as well as many of its variants, one can notice that a line connecting two apparently collinear segments, i.e., segments A and C, across two interrupting parallels looks similar to an optical ray that is refracted when traversing a strip of an optically denser medium. Indeed, the displacement of an optical ray transmitted by a dielectric slab increases with an increase of the angle of incidence and scales with the slab thickness, reproducing thereby the main features of the Poggendorff illusion (Tolansky, 1964). Thus, my hypothesis is that the Poggendorff effect is generated via the accumulated experience of humans with various optical refraction phenomena occurring whenever a rod-like or simply elongated object is immersed in water. One can indeed easily find numerous examples in our everyday life as illustrated in Fig. 2. It is to be noted that very similar pictures can be taken in natural surroundings, and the effects of refraction can easily be enhanced by a bit more carefully arranging the objects to be photographed.

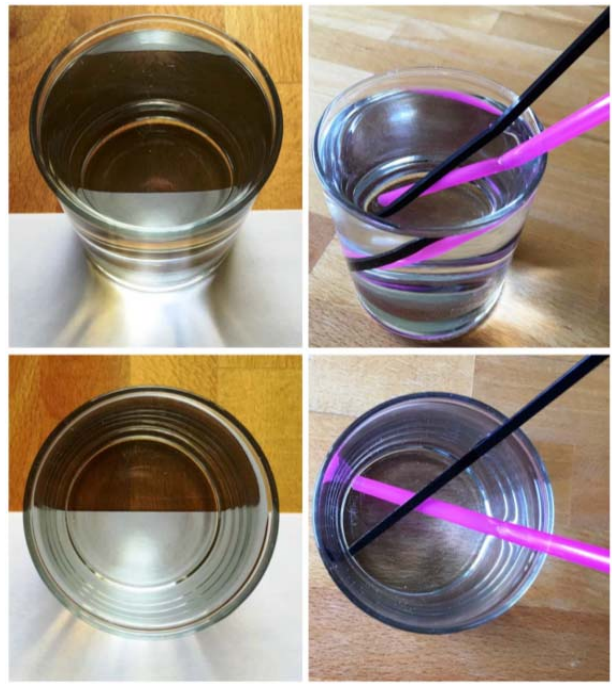

Fig. 2. Illustrations of everyday experience of water-related light refraction phenomena. A straight line (left) or an elongated object (right) appear to be displaced or tilted when seeing through and outside the water (top panels), with the effect disappearing when these objects are viewed perpendicular to the water surfaces (bottom panels).

One advantage of this hypothesis is that the Poggendorff effect, when being associated with it, can straightforwardly be quantified. Indeed, using the law of refraction and elementary trigonometrical operations, one obtains the following expression for the misjudgment $M$ (Fig. 1):

$$M = W\left(\cot\alpha - \frac{\cos\alpha}{\sqrt{N^2 - (\cos\alpha)^2}}\right) , \qquad (1)$$

where $N$ is the effective refractive index in the region between two parallels. I would also consider reasonable to set this effective index somewhere between that of air ($N_a = 1$) and water ($N_w \cong 1.34$), i.e., $N \cong 1.17$, simply because one understands that there is actually no water in considered illustrations akin to Fig. 1. The above relation can greatly be simplified by taking into account the fact that $N - 1 \ll 1$. Assuming also that the acute angle is not too small, one obtains

$$M \cong (N-1)W\cot\alpha , \qquad (2)$$

which transforms into $M \cong 0.17W\cot\alpha$ for $N \cong 1.17$ as argued above. The latter relation, in turn, replicates with the accuracy of 5% the relation mentioned in the introduction and substantiated in (Weintraub & Krantz, 1971). Finally, the above relations for the spatial misjudgement $M$ can easily be converted in the relations for the angular misjudgement $\Delta\alpha$:

$$\Delta\alpha \cong \frac{M}{W}(\sin\alpha)^2 = \frac{\sin 2\alpha}{2}\left(1 - \frac{\sin\alpha}{\sqrt{N^2 - (\cos\alpha)^2}}\right) \cong (N-1)\frac{\sin 2\alpha}{2} . \qquad (3)$$

This relation suggests that the angular misalignment vanishes in the limit of very small angles $\alpha$ as well as for $\alpha \to 90°$, while peaking at $\alpha = 45°$, a feature that has been noted in experimental studies of the Poggendorff effect (Weintraub et al., 1980; Ninio & O'Regan, 1999; Ninio, 2014).

**3. Experiment**

In this experiment, the aim was to establish the statistical properties, primarily the mean value and standard deviation, of the distribution in the effective refractive indexes calculated from the spatial misjudgments obtained for different configurations by using the inversion of Eq. (1):

$$N = \cos\alpha\sqrt{1 + \left(\cot\alpha - \frac{M}{W}\right)^{-2}} . \qquad (4)$$

For example, when treating the misalignment displayed in Fig. 1 using this expression [Eq. (4)], one obtains the effective refractive index $N \cong 1.133$, a value which is quite close to that suggested above for describing the Poggendorff effect.

*3.1. Method*

*3.1.1. Subjects*

The experimental group of seven participants was composed of two 10-year-old children, three engineering students (aged between 20 and 30 years) and two 60-year-old engineers. No special tests of their vision abilities were conducted. It should be emphasized that the tests were carried out in accordance with the Code of Ethics of the World Medical Association (Declaration of Helsinki) and that the informed consent was obtained for conducting these tests.

*3.1.2. Stimuli*

Different configurations, representing vertical or horizontal bars of different widths ($W$ = 30, 40, 50, 55 and 60 mm) that were abutted with parallel segments having different acute angles ($\alpha$ = 30°, 40°, 50°, 60°, and 70°), were displayed (one at a time) on the computer monitor. One of the segments (left or lower) was intentionally displayed far away from the place where it could have been collinear with another (right or upper). This (destined for the adjustment) segment, which is marked with a double-pointed arrow in Fig. 3, was <u>not</u> marked during the experiment.

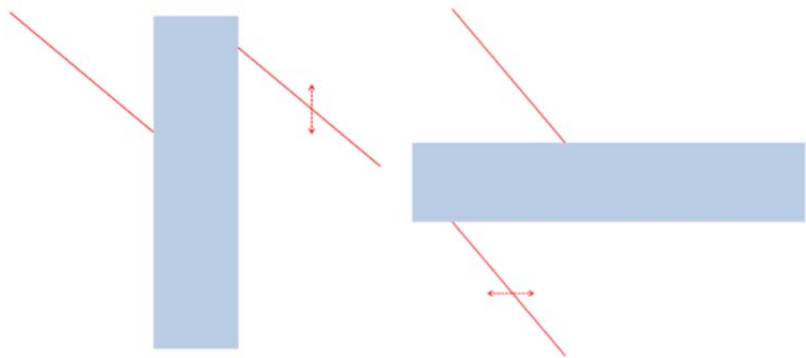

Fig. 3. Examples of two different, vertical (left) and horizontal (right), configurations with a 40-mm-wide bar (when being displayed in full size on a computer monitor) and an acute angle of 40°.

The sequence of configurations consisted of 10 series, first starting with the vertical arrangement progressing for each angle from small to larger widths and then changing to another (progressively decreasing) angle. In this way, the expected misalignment was increasing within each series characterized by the same acute angle, with angles being decreased so that the expected misalignment was again increased. While the prepared sequence of the configuration images was identical, the computer monitors were individual and equipped with personal settings, and no attempts were made to make their appearance identical for all participants.

*3.1.3. Procedure*

Subjects viewed their own computer monitors with the image slides filling the whole screen from their preferable distances. No attempts were made to create special (and identical) conditions and environments for conducting this test. The subjects were instructed to move the off-centred segment using computer keys until it was perceived as collinear with the centred segment. It was also recommended to move the segment back and forth near the "proper" position before deciding on the final placement of the segment. Returns to the previous slides and reconsiderations of the chosen segment position were specifically discouraged. Once the whole series of 50 configurations was completed, it was subjected to the treatment, in which a third long (and parallel to other ones) segment was created and carefully positioned as a continuation of the left or upper segment extending across a bar so that the spatial misjudgement $M$ could be measured. The final data treatment consisted in measurements of widths $W$ and misjudgements $M$ applying an mm-accurate ruler to the computer printouts made on A4-format paper sheets (one sheet per configuration).

3.2. Results

Altogether, the experiment has resulted in 350 sets of different configurations: 175 vertical and 175 horizontal ones characterized by the acute angle $\alpha$, bar width $W$, and the spatial misjudgement $M$. For each set, the effective refractive index was calculated using Eq. (4), and plotted on the same graph (Fig. 4). The results obtained using the observations of the same subject are displayed using the same colour and the same symbol, with filled symbols being assigned to the results obtained with the vertical configurations while open symbols are used only for the horizontal configurations.

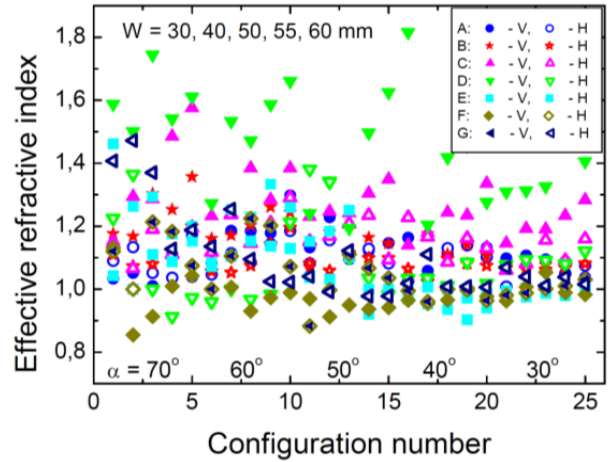

Fig. 4. Effective refractive indexes obtained from the experiments conducted by seven participants marked as A, B, C, D, E, and F. Each data point corresponds to a specific set characterized by the acute angle $\alpha$, bar width $W$, and the spatial misjudgement $M$ in the vertical (filled symbols) or horizontal (open symbols) arrangement. Configuration numbers from 1 to 5 are assigned to the sets with acute angles $\alpha = 70°$ and bar widths $W$ changing from 30 to 60 mm, accordingly. Similarly, configuration numbers from 6 to 10 are assigned to the sets with acute angles $\alpha = 60°$ and bar widths $W$ changing from 30 to 60 mm, accordingly, and so on.

It is interesting and probably important to note that the participant denoted as subject D has consistently arrived at very large misjudgements for the vertical configurations, resulting in high effective indexes, while subject F has repeatedly produced the misjudgements with the opposite sign, resulting in the refractive indexes smaller than unity (Fig. 4). Apart from these apparent deviations, most of the data fill the interval of $1 < N < 1.2$, as can also be seen from the refractive index data histogram (Fig. 5). Tempting though it might be to leave the most divergent data out of the consideration, this approach cannot be justified in our case with so few participants. Statistical treatment of all data (Fig. 4) obtained from this experiment (Fig. 3) resulted in the effective refractive index of the bar $N = 1.13 \pm 0.15$, which is sufficiently close to the average (between that of water and air) refractive index of ~ 1.17 that was suggested to use in Section 2 for describing the Poggendorff effect. Moreover, it is practically identical with the effective refractive index $N \cong 1.133$ obtained from the configuration shown in Fig. 1, replicating Fig. 36 in Tolansky, 1964.

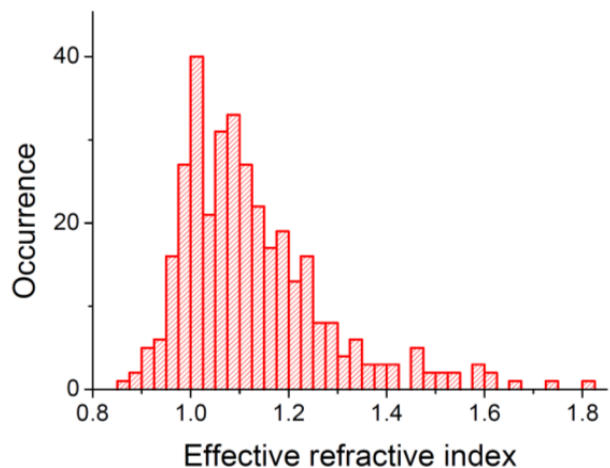

Fig. 5. Histogram of all refractive index data shown in Fig. 4.

## 4. Discussion

The Poggendorff illusion in its classical form of parallel lines interrupting a transversal was considered from the perspective of being related to the everyday experience of observing the light refraction in water. This conceptually novel approach allowed me to immediately obtain the previously reported relationship $M \cong 0.162 W \cot \alpha$ (Weintraub & Krantz, 1971), which relates the spatial misjudgement $M$ to the distance $W$ between the parallels and the acute angle $\alpha$ between the transversal and parallels. It also accounted for the angular misalignment peaking at $\alpha = 45°$, a feature that has been noted in experimental studies of the Poggendorff effect (Weintraub et al., 1980; Ninio & O'Regan, 1999; Ninio, 2014). The experimental results obtained admittedly with a few participants have nonetheless confirmed to a reasonable degree this hypothesis, corroborating the aforementioned relationship, since the majority of the data can be described by the formula $M \cong (1.13 \pm 0.15) W \cot \alpha$. Furthermore, this hypothesis offers a very simple explanation for a particularly puzzling aspect of the Poggendorff illusion that disappears with only acute components being present and re-appears when switching to the obtuse ones (Figs. 6A and 6B). The main difference between these two configurations is, in my opinion, that one has difficulties imagining an occluding strip of an optically dense material, like water, in the former case, while it can be relatively easy done for the latter configuration. Consequently, no refraction experience is invoked in the first case. Finally, both the corner-Poggendorff illusion and the Hering illusion (Figs. 6C and 6D) can be viewed and explained from the perspective of light refraction by a prism, when a light ray propagating through a prism is deflected towards a prism base (i.e., away from a prism top).

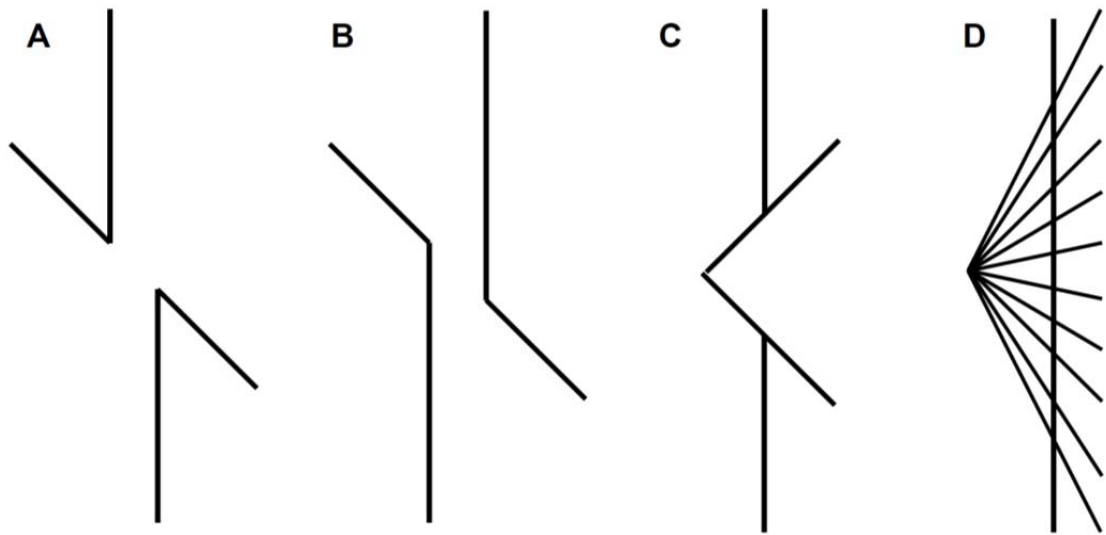

Fig. 6. Other geometrical illusions that can be related to observing the light refraction in water. The illusion is abolished with only acute components of the Poggendorff pattern being present (A), but restored when switching to the obtuse components (B), because it is very difficult to imagine the presence of an occluding strip (with a higher refractive index) in the former and very easy in the latter case. The corner-Poggendorff illusion (C) and the Hering illusion (D) can both be related to the well-known optical refraction phenomenon, viz., a light ray propagating through a prism is deflected towards a prism base (i.e., away from a prism top).

## 5. Conclusions

This work is devoted to presenting a conceptually novel explanation for one of the most prominent optical illusion, the Poggendorff illusion. It was argued that this illusion in its classical form of parallel lines interrupting a transversal can be viewed from the perspective of being related to the everyday experience of observing the light refraction in water. The main idea is that the space between two parallels is perceived as an optically denser (than air) medium cause refraction of the transversal ray. This approach allowed me to consistently account, both qualitatively and quantitatively, for most of the features associated with the Poggendorf illusion. Statistical treatment of the visual experiments conducted with 7 participants, each analysing 50 configurations having different intercepting angles and strip widths, resulted in the effective refractive index of the occluding strip $N = 1.13 \pm 0.15$, which is sufficiently close to the average (between that of water and air) refractive index of ~ 1.17. Finally, I have further argued that the same mechanism can also be employed to account for many variants of the Poggendorf illusion, including the corner-Poggendorf pattern, as well as for the Hering illusion. Overall, in my opinion, there are sufficiently solid indications in favour of my explanation of the Poggendorff illusion.

**Acknowledgements**

The author thanks voluntary participants for taking part in the described experiment, with special thanks to Elena Bozhevolnaya for the support and invaluable criticism that made writing this account possible. Finally, I am very grateful to my father who initiated my interest in physics, in general, and in optical phenomena, in particular, persuading to always seek exciting effects and to follow one's own curiosity.

# References

Day, R.H. & Dickinson, R. G. (1976). The components of the Poggendorff illusion. *Br. J. Psychol.* 67, 537–552.

Gillam, B. (1971). A depth processing theory of the Poggendorff illusion. *Percept. Psychophys*. 10, 211–215.

Howe, C. Q., Yang, Z., & Purves, D. (2005). The Poggendorff illusion explained by natural scene geometry. *Proc. Nat. Acad. Sci. U.S.A.* 102, 7707–7712.

Ninio, J. & O'Regan, J. K. (1999).Characterization of the misalignment and misangulation components in the Poggendorff and corner-Poggendorff illusions. *Perception* 28, 949–964.

Ninio, J. (2014). Geometrical illusions are not always where you think they are: a review of some classical and less classical illusions, and ways to describe them. *Frontiers in Human Neuroscience*, 8, art.No.856 (21).

Phillips, D. (2006). The Poggendorff illusion: premeditated or unpremeditated misbehaviour? *Perception* 35, 1709–1712.

Pressey, A.W. & Sweeney, O. (1972). Acute angles and the Poggendorff illusion. *Q. J. Exp. Psychol.* 24, 169–174.

Talasli, U. & Inan, A. B. (2015). Applying Emmert's law to the Poggendorff illusion. *Frontiers in Human Neuroscience*, 9, art.No.531 (12).

Tolansky, S. (1964). *Optical illusions*. Oxford: Pergamon Press.

Weintraub, D. J., & Krantz, D. H. (1971). The Poggendorff illusion: Amputations, rotations, and other perturbations. *Perception & Psychophysics*, 10, 257-264.

Weintraub, D. J. & Tong, L. (1974). Assessing Poggendorff effects via collinearity, perpendicularity, parallelism and Oppel (distance) experiments. *Percept. Psychophys.* 16, 213–221.

Weintraub, D. J., Krantz, D. H., & Olson, T. P. (1980).The Poggendorff illusion: consider all the angles. *J. Exp. Psychol. Hum. Percept. Perform.* 6, 718–725.